\documentclass[aps,prl,footinbib,twocolumn,superscriptaddress]{revtex4-1}
\usepackage{amsmath}
\usepackage{amssymb}
\usepackage{epsfig}
\usepackage{graphics}
\usepackage{bm}
\usepackage{xcolor}

\usepackage{tikz}
\usetikzlibrary{positioning,arrows}
\usetikzlibrary{decorations.pathmorphing}
\usetikzlibrary{decorations.markings}

\usetikzlibrary{calc}

\tikzset{
  fermion/.style={draw=black,postaction={decorate}, thick,
    decoration={markings,mark=at position #1 with {\arrow[black,scale=.6]{triangle 45}}}},
  inoutbec/.style={draw=black,postaction={decorate}, thick, densely dotted,
    decoration={markings,mark=at position #1 with {\arrow[black,scale=.6]{triangle 45}}}},
  bec/.style={draw=black,postaction={decorate}, thick, dashed,
    decoration={markings,mark=at position #1 with {\arrow[black,scale=.6]{triangle 45}}}},
  Tmatrix/.style={rectangle, solid, draw, fill=lightgray, minimum width = 25pt, minimum height = 25pt},
  photon/.style={decorate, draw=black, decoration={coil,aspect=0}, thick},
}

\newcommand{\beq}{\begin{equation}}
\newcommand{\eeq}{\end{equation}}
\newcommand{\bea}{\begin{eqnarray}}
\newcommand{\eea}{\end{eqnarray}}

\def\bk{{\bf k}}
\def\bp{{\bf p}}
\def\bq{{\bf q}}

\def\bP{{\bf P}}

\def\calV{\mathcal{V}}

\def\calT{\mathcal{T}}

\def\nn{\nonumber}

\def\Re{\mathfrak{Re}}
\def\Im{\mathfrak{Im}}

\begin{document}
\title{A mixed dimensional Bose polaron}
\author{Niels Jakob S\o e Loft}
\affiliation{Department of Physics and Astronomy,  Aarhus University, Ny Munkegade, DK-8000 Aarhus C, Denmark}
\author{Zhigang Wu}
\affiliation{Institute for Advanced Study, Tsinghua University, Beijing, 100084, China}
\author{G.\ M.\ Bruun}
\affiliation{Department of Physics and Astronomy,  Aarhus University, Ny Munkegade, DK-8000 Aarhus C, Denmark}

\date{\today}
\begin{abstract}
  A new generation of cold atom experiments trapping atomic mixtures
  in species selective optical potentials opens up the intriguing
  possibility to create systems in which different atoms live in
  different spatial dimensions. Inspired by this, we investigate a
  mixed dimensional Bose polaron consisting of an impurity particle
  moving in a two-dimensional (2D) layer immersed in a 3D
  Bose-Einstein condensate (BEC), using a theory that includes the
  mixed dimensional vacuum scattering between the impurity and the
  bosons exactly. We show that similarly to the pure 3D case, this
  system exhibits a well-defined polaron state for attractive
  boson-impurity interaction that evolves smoothly into a
  mixed-dimensional dimer for strong attraction, as well as a
  well-defined polaron state for weak repulsive interaction, which
  becomes over-damped for strong interaction.  We furthermore find
  that the properties of the polaron depend only weakly on the gas
  parameter of the BEC as long as the Bogoliubov theory remains a
  valid description for the BEC. This indicates that higher order
  correlations between the impurity and the bosons are suppressed by
  the mixed dimensional geometry in comparison to a pure 3D system,
  and that the mixed dimensional polaron has universal properties in
  the unitarity limit of the impurity-boson interaction.
\end{abstract}
\pacs{$\ldots$}
\maketitle
\section{introduction}
The problem of a mobile impurity particle in a quantum reservoir plays
a central role in physical systems across many energy scales, ranging
from $^3$He-$^4$He mixtures~\cite{BaymPethick1991book} and polarons in
condensed matter systems~\cite{LandauPekar,Mahan2000book}, to
elementary particles surrounded by the Higgs field giving them their
mass~\cite{Weinberg1995}.  Our understanding of the impurity physics
has improved significantly with the experimental realisation of highly
population-imbalanced atomic gases, where the minority atoms play the
role of the impurities, and the majority atoms constitute the quantum
environment.  A powerful feature of atomic gases is that the
interaction between the impurity atom and the surrounding gas can be
tuned experimentally using Feshbach resonances~\cite{Chin2010}. This
opens up the possibility to systematically study the effects of strong
correlations between the impurity and the environment. The first
experiments realised impurity atoms in a degenerate Fermi
gas~\cite{Schirotzek2009,Kohstall2012,Koschorreck2012}, coined the
Fermi polaron, for which we now have accurate theories even in the
case of strong
interactions~\cite{Chevy2006,Prokofev2008,Mora2009,Punk2009,Combescot2009,Cui2010,Massignan2011,Massignan_Zaccanti_Bruun,Cui2015}.
The Bose polaron, i.e.\ an impurity atom in a Bose-Einstein condensate
(BEC), has been realised in a 1D geometry~\cite{Catani2012} as well as
in 3D~\cite{Jorgensen2016,Hu2016}. Whereas early theories for the Bose
polaron were based on the so-called Fr\"ohlich
model~\cite{Cucchietti2006,Huang2009,Tempere2009,Grusdt2015},
perturbation theory explicitly shows that this model breaks down at
third order in the interaction strength~\cite{Christensen2015}. Using
a microscopic theory, the results of a diagrammatic
calculation~\cite{Rath2013}, a variational ansatz including Efimov
physics~\cite{Levinsen2015} or the dressing by many Bogoliubov
modes~\cite{Shchadilova2016}, and Monte-Carlo
calculations~\cite{Ardila2015,Ardila2016} all give results consistent with the
experimental data. Most recently, it was shown that even Efimov
physics can be detected in the Bose polaron spectrum~\cite{Sun2017}.

An exciting development is the creation of novel mixed dimensional
systems using cold atoms in species selective optical
lattices~\cite{Lamporesi2010,McKay2013,Jotzu2015,LeBlanc2007,Catani2009,Haller2010}. The
mixed dimensional geometry gives rise to new effects already at the
few-body level such as a strong enhancement of the interaction between
atoms by confinement induced resonances~\cite{Nishida2008}. At the
many-body level, these systems have been predicted to give rise to a
plethora of interesting phenomena including strong induced
interactions~\cite{Suchet2017}, enhanced Kondo
coupling~\cite{Cheng2017} and unconventional superfluid
phases~\cite{Iskin2010,Okamoto2017}, some of which with nontrivial
topological
properties~\cite{Nishida2009,Kim2013,Wu2017,Midtgaard2017}. Since the
polaron problem has proven to be a powerful probe into strong
correlations, it is of interest to examine this problem in a mixed
dimensional setup.

\begin{figure}[htbp]
  \centering
  \includegraphics[width=\columnwidth]{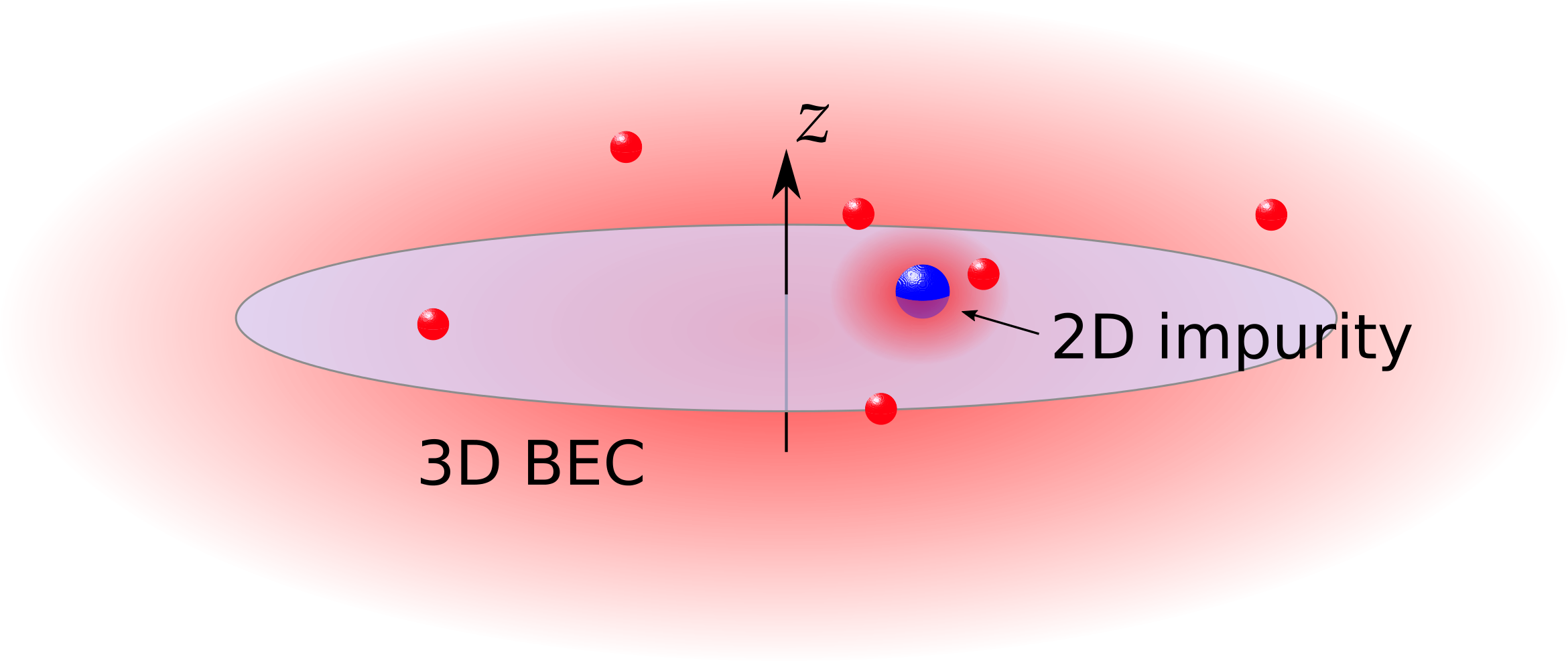}
  \caption{Sketch of the system: 2D impurity particle (blue) immersed in a 3D Bose-Einstein condensate (red).}
  \label{fig:sketch}
\end{figure}

In this paper, we examine a mixed dimensional Bose polaron, where the
impurity particle is confined to move in a 2D plane immersed in a 3D
BEC (see Fig.~\ref{fig:sketch}).  Using a diagrammatic ladder
approximation, which includes the mixed dimensional 2D-3D vacuum
scattering between the impurity and the bosons exactly, we calculate
the quasiparticle properties of the polaron as a function of the
impurity-boson interaction strength and the gas parameter of the
BEC. We show that the impurity problem has the same qualitative
features as that for the pure 3D case. There is a well-defined
quasiparticle for attractive impurity-boson interaction (attractive
polaron), which smoothly evolves into a mixed-dimensional dimer state
consisting of a boson in 3D bound to the impurity in the plane for
strong interaction. For repulsive impurity-boson interaction, there is
also a well-defined quasiparticle state (repulsive polaron), which
becomes over-damped for strong interaction. The theory predicts that
the dependence of the properties of the polaron on the gas parameter
of the BEC is weaker than in the pure 3D case.  This indicates that
the polaron has universal properties in the unitarity limit of the
impurity-boson interaction. We argue that this could be due to the
fact that the effects of the impurity on the bosons are limited by the
mixed dimensional geometry such that higher order correlations are
suppressed.

\section{Model}
We consider a single impurity atom of mass $m$ confined in the 2D
$xy$-plane by a strong harmonic trap $m\omega_z^2z^2/2$ along the
$z$-direction. Since only one impurity is considered, our results of
course do not depend on the statistics of the impurity.  For
concreteness, we take the impurity to be a fermion. The impurity atom
is immersed in a weakly interacting 3D Bose gas of atoms with mass
$m_B$ (see Fig.~\ref{fig:sketch}). The bosons form a BEC with density
$n_0$, which is accurately described by Bogoliubov theory since we
assume $n_0^{1/3} a_{B} \ll 1$, where $a_B$ is the boson scattering
length. The Hamiltonian of the system is
\begin{equation}
  \label{eq:FullH}
  H = \sum_{\bp_\perp} \frac{\bp_\perp^2}{2m}
  a_{\bp_\perp}^\dagger a_{\bp_\perp}
  + \sum_{\bp} E_\bp \gamma_\bp^\dagger \gamma_\bp
  + H_\text{int} \; ,
\end{equation}
where $a_{\bp_\perp}^\dagger$ creates an impurity with 2D momentum
$\bp_\perp=(p_x,p_y)$, and $\gamma_\bp^\dagger$ creates Bogoliubov
mode in the BEC with 3D momentum $\bp$ and energy $E_\bp =
\sqrt{\epsilon_\bp (\epsilon_\bp + 2n_B g_B)}$. Here $\epsilon_\bp =
\bp^2 / 2m_B$ and $g_B=4\pi a_B/m_B$.  Throughout this paper, we set
$\hbar = k_B = 1$. For clarity we will use the $\perp$ sign to denote
 vectors in the plane in order to distinguish them from the 3D
vectors. The interaction between the bosons and the impurity is
\begin{equation}
  \label{eq:Hint2}
  H_\text{int}
  = \frac{1}{\calV}
  \sum_{\bp \bp'_\perp \bq}
  e^{-(q_z l_z / 2)^2} V(\bq) b^\dag_{\bp+\bq}   a^\dag_{\bp'_\perp-\bq_\perp}  a_{\bp'_\perp} b_{\bp} \; ,
\end{equation}
where $\bq = (\bq_\perp, q_z)$, $l_z=1/\sqrt{m\omega_z}$ is the
harmonic oscillator length for the vertical trap and $V(\bq)$ is the
boson-fermion interaction potential. The latter will later be
eliminated in favor of the effective 2D-3D scattering length $a_{\rm
  eff}$. The operator $b_\bp^\dagger$ creates a boson with momentum
$\bp$, and it is related to the Bogoliubov mode creation operators by
the usual relation $b_\bp = u_\bp \gamma_\bp -
v_\bp\gamma^\dagger_{-\bp}$ with $u_\bp^2 = [(\epsilon_\bp +
g_{B}n_B)/E_\bp + 1]/2$ and $v_\bp^2 = [(\epsilon_\bp +
g_{B}n_B)/E_\bp - 1]/2$.  We have in Eq.~\eqref{eq:Hint2} assumed that
due to the strong confinement, the impurity resides in the lowest
harmonic oscillator state $\phi_0(z)=\exp\left
  \{-z^2/2l_z^2\right\}/\pi^{1/4}\sqrt{l_z}$ in the $z$-direction. The
exponential factor in Eq.\ (\ref{eq:Hint2}) comes from the Fourier
transform of $\phi_0(z)$.  Note that only transverse momentum is
conserved during boson-impurity collisions due to the confinement of
the impurity in the vertical direction.

\section{Self-energy}
We employ the ladder approximation to calculate the self-energy of the
Bose polaron~\cite{Rath2013}. For the Fermi polaron, this
approximation has proven to be surprisingly accurate even for strong
interactions~\cite{Chevy2006,Prokofev2008,Mora2009,Punk2009,Combescot2009,Massignan2011,Massignan_Zaccanti_Bruun}.
The accuracy of the ladder approximation is less clear for the Bose
polaron since there is no Pauli principle, which suppresses more than
one fermion from being close to the impurity. The ladder approximation
neglects such higher order correlations, which for instance can lead
to the formation of a 3-body Efimov state consisting of the impurity
atom and two bosons. In Ref.~\cite{Levinsen2015}, it was shown that
these Efimov correlations are important when the scattering length
$a_-$ for which the first Efimov trimer occurs, is comparable to or
smaller than the interparticle distance in the BEC, whereas their
effects are small for larger $a_-$. It has also been shown that the
Efimov effect is suppressed in reduced dimensions as compared to the
pure 3D case~\cite{Nishida2011}.  We therefore assume that higher
order correlations are suppressed in the mixed dimensional geometry,
and we resort to the ladder approximation in the following.

\begin{figure}[htbp]
  \centering

  \begin{tikzpicture}
    \def \dist {1.2};

    
    \node (a) {a)};
    \coordinate [right = \dist of a] (center);
    \coordinate [left = .3 of center] (aux1);
    \coordinate [left = .8 of aux1] (e1);
    \coordinate [right = .3 of center] (aux2);
    \coordinate [right = .8 of aux2] (e2);
    \draw [fermion=.6] (e1) -- (aux1);
    \draw [fermion=.7] (aux2) -- (e2);
    \node (sigma) [circle, solid, draw, fill=lightgray,
    minimum height = 25pt] at (center) {$\Sigma$};
    \node [above = of center] {$\Sigma$};

    \node (s1) [right = \dist of center] {$=$};
    \node [above = .9 of s1] {$=$};

    \coordinate [right = \dist of s1] (center);
    \node (interaction) [Tmatrix] at (center) {$\mathcal{T}$};
    \coordinate [below left = .5 of interaction.south west] (e1);
    \coordinate [below right = .5 of interaction.south east] (e2);
    \coordinate [above left = .5 of interaction.north west] (f1);
    \coordinate [above right = .5 of interaction.north east] (f2);    
    \draw [fermion=.75] (e1) -- (interaction.south west);
    \draw [fermion=.75] (interaction.south east) -- (e2);
    \draw [inoutbec=.75] (f1) -- (interaction.north west);
    \draw [inoutbec=.75] (interaction.north east) -- (f2);
    \node [above = of center] {$\Sigma_0$};
    
    \node (s1) [right = of center] {$+$};
    \node [above = .85 of s1] {$+$};

    \coordinate [right = \dist of s1] (center);
    \node (interaction) [Tmatrix] at (center) {}; 
    \draw [bec=.31] ($(interaction.north west)!.5!(interaction.north east)$) circle (12.5pt);
   \node (interaction) [Tmatrix] at (center) {$\mathcal{T}$};
    \coordinate [below left = .5 of interaction.south west] (e1);
    \coordinate [below right = .5 of interaction.south east] (e2);
    \draw [fermion=.75] (e1) -- (interaction.south west);
    \draw [fermion=.75] (interaction.south east) -- (e2);
    \node [above = of center] {$\Sigma_1$};


    \node (b) [below = \dist of a] {b)};

    \coordinate [right = \dist of b] (center);
    \node (interaction) [Tmatrix] at (center) {$\mathcal{T}$};
    \coordinate [below left = .5 of interaction.south west] (e1);
    \coordinate [below right = .5 of interaction.south east] (e2);
    \coordinate [above left = .5 of interaction.north west] (f1);
    \coordinate [above right = .5 of interaction.north east] (f2);    
    \draw [fermion=.75] (e1) -- (interaction.south west);
    \draw [fermion=.75] (interaction.south east) -- (e2);
    \draw [bec=.75] (f1) -- (interaction.north west);
    \draw [bec=.75] (interaction.north east) -- (f2);

    \node (s1) [right = \dist of center] {$=$};

    \coordinate [right = .8*\dist of s1] (center);
    \coordinate [above = 12.5pt of center] (aux1);
    \coordinate [below = 12.5pt of center] (aux2);
    \coordinate [below left = .5 of aux2] (e1);
    \coordinate [below right = .5 of aux2] (e2);
    \coordinate [above left = .5 of aux1] (f1);
    \coordinate [above right = .5 of aux1] (f2);
    \node [left] at (center) {$g$};
    \draw [photon] (aux1) -- (aux2);
    \draw [fermion=.75] (e1) -- (aux2);
    \draw [fermion=.75] (aux2) -- (e2);
    \draw [bec=.75] (f1) -- (aux1);
    \draw [bec=.75] (aux1) -- (f2);

    \node (s1) [right = .6*\dist of center] {$+$};

    \coordinate [right = .8*\dist of s1] (center1);
    \coordinate [above = 12.5pt of center1] (aux1);
    \coordinate [below = 12.5pt of center1] (aux2);
    \coordinate [below left = .5 of aux2] (e1);
    \coordinate [above left = .5 of aux1] (f1);
    \coordinate [right = .8*\dist of center1] (center2);
    \node (interaction) [Tmatrix] at (center2) {$\mathcal{T}$};
    \coordinate [below right = .5 of interaction.south east] (e2);
    \coordinate [above right = .5 of interaction.north east] (f2);
    \node [left] at (center1) {$g$};
    \draw [photon] (aux1) -- (aux2);
    \draw [fermion=.75] (e1) -- (aux2);
    \draw [fermion=.7] (aux2) -- (interaction.south west);
    \draw [fermion=.75] (interaction.south east) -- (e2);
    \draw [bec=.75] (f1) -- (aux1);
    \draw [bec=.7] (aux1) -- (interaction.north west);
    \draw [bec=.75] (interaction.north east) -- (f2);

  \end{tikzpicture}

  \caption{Diagrams used in the ladder approximation for the impurity. A solid
    line represents an impurity  propagator, a dashed line denotes a
    boson propagator, and a dotted line denotes a boson emitted or absorbed by the BEC.
     {\bf a)} The polaron self-energy given by the
    sum of the diagrams $\Sigma_0$ and $\Sigma_1$. {\bf b)} The
    $\calT$-matrix giving the scattering  between the impurity  and a boson. 
    }
  \label{fig:Diagrams}
\end{figure}
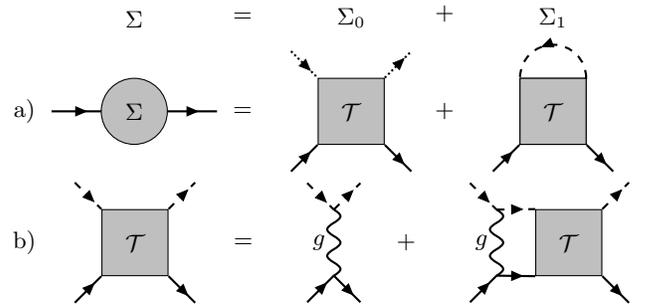

Within the ladder approximation, the polaron self-energy for
momentum-frequency $(\bk_\perp, i\omega_n)$ is given by (see Fig.\
\ref{fig:Diagrams}a)
\begin{equation}
  \label{eq:Self-Energy}
  \Sigma(\bk_\perp,  i\omega_n)
  = \Sigma_0(\bk_\perp,  i\omega_n) + \Sigma_1(\bk_\perp,  i\omega_n)
\end{equation}
where
\begin{align}
  \label{eq:Diagram1}
  \Sigma_0(\bk_\perp,i\omega_n) =
  n_B \calT(\bk_\perp,i\omega_n) 
  \end{align}
  describes the scattering of bosons out of the condensate by the
  impurity. Here $\omega_n = (2n+1)\pi T$ is a fermionic Matsubara
  frequency where $T$ is the temperature and $n$ is an integer, and $ \calT$ is the mixed dimension scattering matrix (see below).  The
self-energy coming from the  scattering of bosons not in the condensate is
\begin{align}
  \Sigma_1(\bk_\perp,i\omega_n) &= - T\sum_{\nu}\int\! \frac{d^3p}{(2\pi)^3} G_{ 11}(\bp,i\omega_\nu) \nn \\
  &\qquad\qquad\quad\times\calT(\bk_\perp + \bp_\perp,i\omega_n+i\omega_\nu),
 \label{eq:Diagram2}
 \end{align}
 where $\omega_\nu = 2\nu \pi T$ is a bosonic Matsubara frequency with
 $\nu$ being an integer. The normal Bogoliubov Green's function for
 the bosons is
\begin{equation}
  \label{eq:GB11}
  G_{11}(\bq, i\omega_\nu) =
  \frac{u_\bq^2}{i\omega_\nu - E_\bq}
  - \frac{v_\bq^2}{i\omega_\nu + E_\bq} \; .
\end{equation}
The 2D-3D scattering matrix between the impurity and a boson can be written as
(see Fig.\ \ref{fig:Diagrams}b)~\cite{Nishida2009}
\begin{equation}
  \label{eq:T-matrix}
  \calT (\bP_\perp,i\omega_m) = \frac{1}{g^{-1} -
    \Pi(\bP_\perp,i\omega_m)} \; .
\end{equation}
Here $g = 2\pi a_\text{eff} / \sqrt{m_Bm_r}$, $m_r=mm_B/(m+m_B)$ is
the reduced mass, $a_\text{eff}$ is the effective 2D-3D scattering
length and $\Pi(\bP_\perp,i\omega_n)$ is the pair propagator. The
effective scattering length is a function of the 3D boson-impurity
scattering length and the trap harmonic oscillator length $l_z$ along
the $z$-direction. This leads to several confinement induced
resonances, which can be exploited to tune the  2D-3D
interaction strength~\cite{Nishida2008}.

The mixed-dimensional pair propagator is given by
\begin{align}
  \Pi(\bP_\perp,i\omega_m) = &
  -T\sum_{\nu}\int\! \frac{d^3p}{(2\pi)^3}
  G_{11}(\bp,i\omega_\nu) \nn \\
 &\qquad\quad \times 
  G^0(\bP_\perp - \bp_\perp,i\omega_m-i\omega_\nu) \; ,
\end{align}
where $G^0(\bq, i\omega_n) = [i\omega_n - \xi_\bq]^{-1}$ is the
bare impurity propagator with $\xi_\bq = \bq^2/2m - \mu$ the bare
energy relative to the impurity chemical potential, $\mu$, taken to
minus infinity since there is only a single impurity. Performing the
Matsubara summation we arrive at
   \begin{align}
    \Pi(\bP_\perp,i\omega_m)&= \int \frac{d^3p}{(2\pi)^3} \left[
   \frac{u_\bp^2(1 + f_\bp) }{i\omega_m - E_\bp - \xi_{\bP_\perp - \bp_\perp}}\right. \nn
   \\ 
  + &\left. \frac{v_\bp^2f_\bp }{i\omega_m + E_\bp - \xi_{\bP_\perp - \bp_\perp}}
  + \frac{2m_B}{p^2 +  p_\perp^2/\alpha}
  \right]  ,
    \label{eq:PairProp1}
\end{align}
where $f_\bp = [\exp(E_\bp/T) - 1]^{-1}$ is the Bose distribution
function and $\alpha=m/m_B$ is the ratio of the impurity and boson
masses.  The last term in the brackets in Eq.~\eqref{eq:PairProp1}
comes from the regularization of the pair propagator by identifying
the molecular pole of the $\calT$-matrix at zero center-of-mass
momentum in vacuum with $\omega_M=-1/2m_ra_\text{eff}^2$ for
$a_\text{eff}>0$~\cite{Nishida2009}.

Equations (\ref{eq:Self-Energy})-(\ref{eq:PairProp1}) have the usual
structure of the ladder approximation for a 3D Fermi polaron apart
from two differences: First, the scattering medium is a BEC which involves
processes describing the scattering of bosons into and out of the
condensate; second, the mixed dimension 2D-3D scattering geometry has
no intrinsic rotational symmetry, which complicates the evaluation of
the resulting integrals significantly compared to the usual 3D case,
as we shall discuss below.
  
\section{Quasiparticle properties}
The quasiparticle properties of the mixed dimension polaron are
encapsulated in the single-particle retarded Green's function
\begin{align}
G(\bk_\perp,\omega) = \frac{1} {\omega+ i0^+ - \Sigma(\bk_\perp,\omega) },
\end{align}
where $\Sigma(\bk_\perp,\omega)$ is the retarded polaron self-energy
obtained from performing the analytical continuation $i\omega_n + \mu
\rightarrow \omega + i0^+$.  To characterize the quasiparticle, we
calculate its dispersion, residue, and effective mass. The
quasiparticle dispersion $\varepsilon_{\bk_\perp}$ for a given
momentum $\bk_\perp$ is found by solving the self-consistent equation
\begin{equation}
  \label{eq:Energy}
  \varepsilon_{\bk_\perp} =\frac{\bk_\perp^2}{2m}+ \Re\Sigma(\bk_\perp, \varepsilon_{\bk_\perp})  \; ,
\end{equation}
where we assume that the damping (determined by the imaginary part of
$\Sigma$) of the polaron is small. The quasiparticle residue is
\begin{equation}
  \label{eq:SpectralWeight}
  Z_{\bk_\perp} = \frac{1}{1 -
    \partial_\omega \Re\Sigma(\bk_\perp, \omega)|_{\omega=\varepsilon_{\bk_\perp}}}
  \; ,
\end{equation}
and the effective mass is
\begin{equation}
  \label{eq:Mass}
  m^*_{\bk_\perp} = \frac{Z_{\bk_\perp}^{-1}}{m^{-1} +
    k_\perp^{-1} \partial_{k_\perp} \Re\Sigma(\bk_\perp, \omega)
    |_{\omega=\varepsilon_{\bk_\perp}}} \; .
\end{equation}
It should be noted that $\Sigma$ only depends on the length of
$\bk_\perp$, denoted $k_\perp$ above. We shall also calculate the spectral
function of the polaron defined as
\begin{align}
A(\bk_\perp,\omega)=-2\Im G(\bk_\perp,\omega) \; .
\label{eq:SpectralFct}
\end{align}

\section{Numerical calculation}
The mixed dimensional geometry turns out to significantly complicate
the numerical calculation of the polaron self-energy. The reason is
that the scattering of the impurity on a boson does not conserve
momentum along the $z$-direction and therefore has no rotational
symmetry, which can be used to reduce the number of
convoluted integrals in the self-energy. This means that in order to
make progress, we have to use simplifications for the calculation of
$\Sigma_1(\bk_\perp,\omega)$ given by Eq.~(\ref{eq:Diagram2}), which
involves six convoluted integrals. For $\Sigma_1(\bk_\perp,\omega)$ we
shall approximate the mixed dimension pair propagator by that for a
non-interacting Bose gas. Since we focus on the case of zero
temperature, the pair propagator is then given by the vacuum
expression
\begin{align}
\Pi_{\rm vac}(\bP_\perp, i\omega_m) = -i\,
\frac{\sqrt{m_B}m_r}{\sqrt2\pi}
\sqrt{i\omega_m + \mu -\frac{\bP_\perp^2}{2M}} \; ,
\label{eq:PairpropagatorVacuumAna}
\end{align}
where $M = m+m_B$ and the complex square root is taken in the upper
half plane. Physically, this approximation corresponds to
assuming that the boson-impurity scattering is unaffected by the BEC
medium, which is a good approximation for momenta $p\gtrsim 1/\xi_B$,
where $\xi_B=1/\sqrt{8\pi n_0a_B}$ is the coherence length of the BEC.
With this approximation, the numerical evaluation of
$\Sigma_1(\bk_\perp, \omega)$ becomes feasible. In the following, we
shall suppress the momentum label $\bk_\perp$ for the polaron, as we
only consider the case of a zero momentum polaron $\bk_\perp =
\bf{0}$. We refer the reader to the appendix for details of the
numerical procedure.

\section{Results}
\begin{figure}[htbp]
  \centering
  \includegraphics[width=1.07\columnwidth]{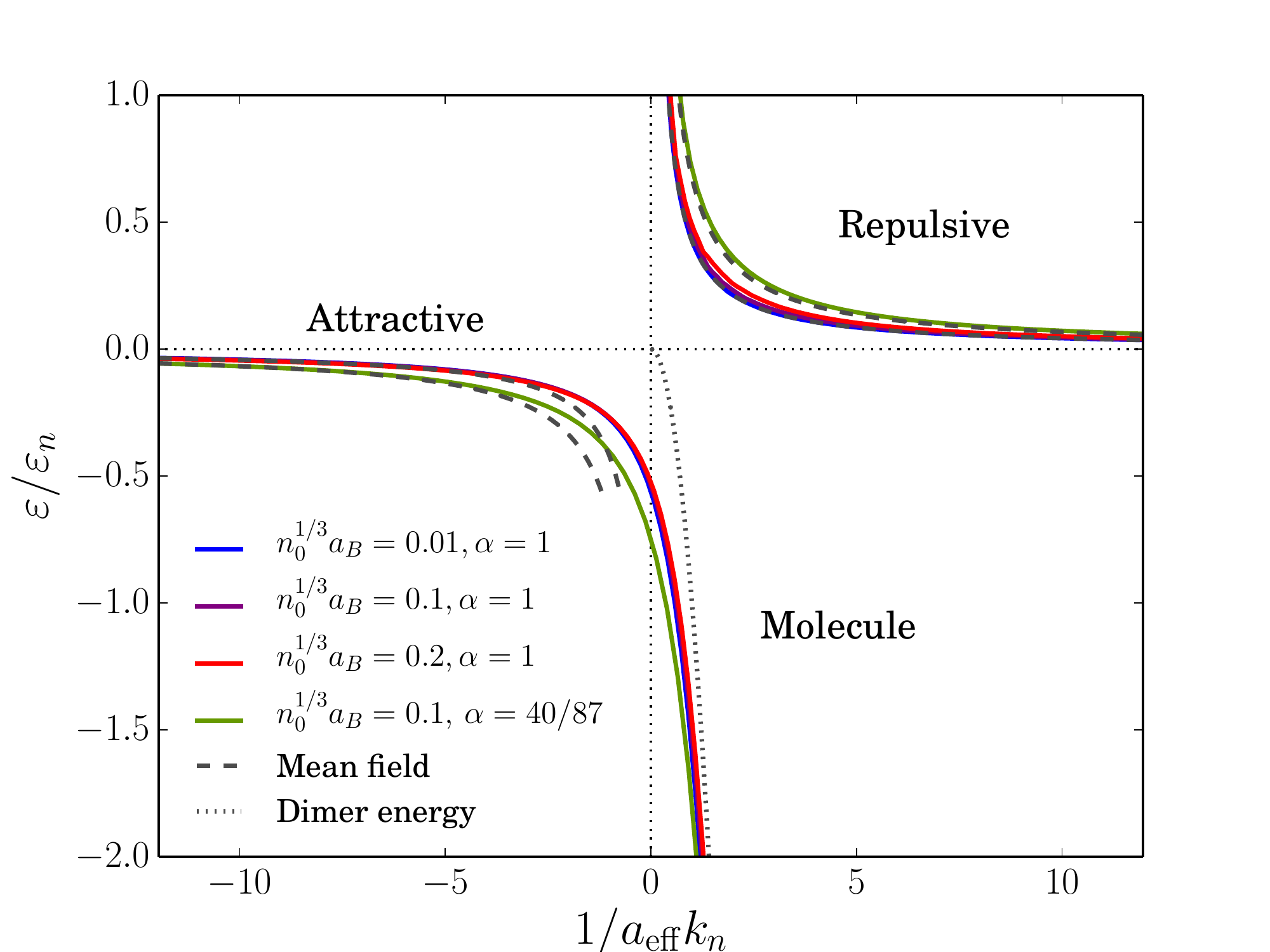}
  \caption{The quasi-particle energy for zero momentum as a function
    of the inverse Fermi-Bose interaction strength.}
  \label{fig:Energy}
\end{figure}

In this section, we present numerical results for the quasi-particle
properties of the Bose polaron.  In Fig.\ \ref{fig:Energy}, we plot
the polaron energy $\varepsilon/\varepsilon_n$ for zero momentum as a
function of the inverse coupling strength $1/k_na_\text{eff}$ at zero
temperature.  We have defined the momentum and energy scales as
$k_n=(6\pi^2 n_B)^{1/3}$ and $\varepsilon_n=k_n^2/2m_B$
respectively. The energy is calculated for various gas parameters
$n_0^{1/3}a_B$ of the BEC, and for the mass ratios $\alpha=m/m_B=1$
and $\alpha=1,40/87$ relevant for the experiments in Ref.\
\cite{Jorgensen2016} and \cite{Hu2016}.
\begin{figure}[htbp]
  \centering
  \includegraphics[width=1.07\columnwidth]{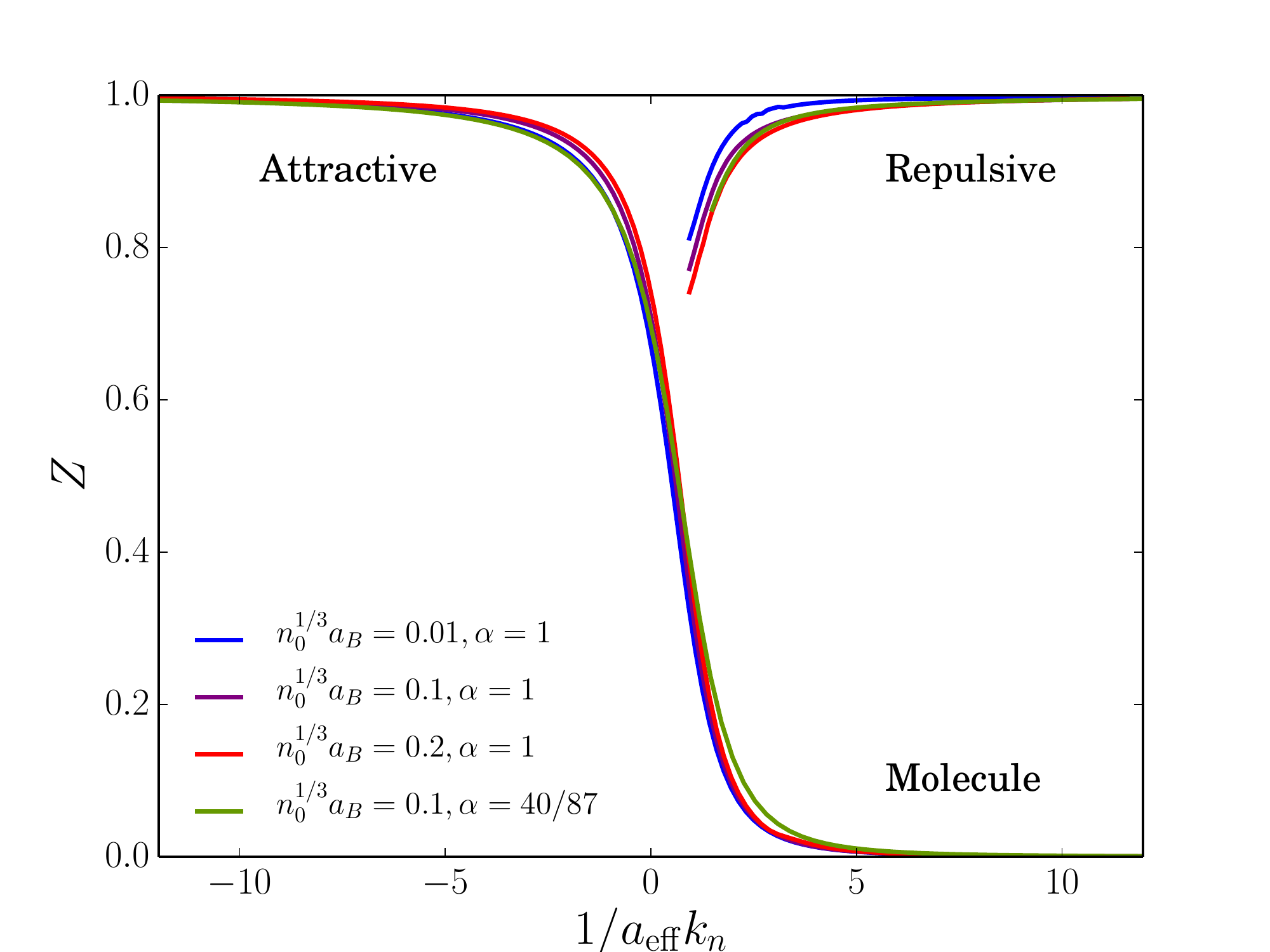}
  \caption{The quasiparticle residue for zero momentum as a function
    of the inverse Fermi-Bose interaction strength.}
  \label{fig:Spectral}
\end{figure}
\begin{figure}[htbp]
  \centering
  \includegraphics[width=1.07\columnwidth]{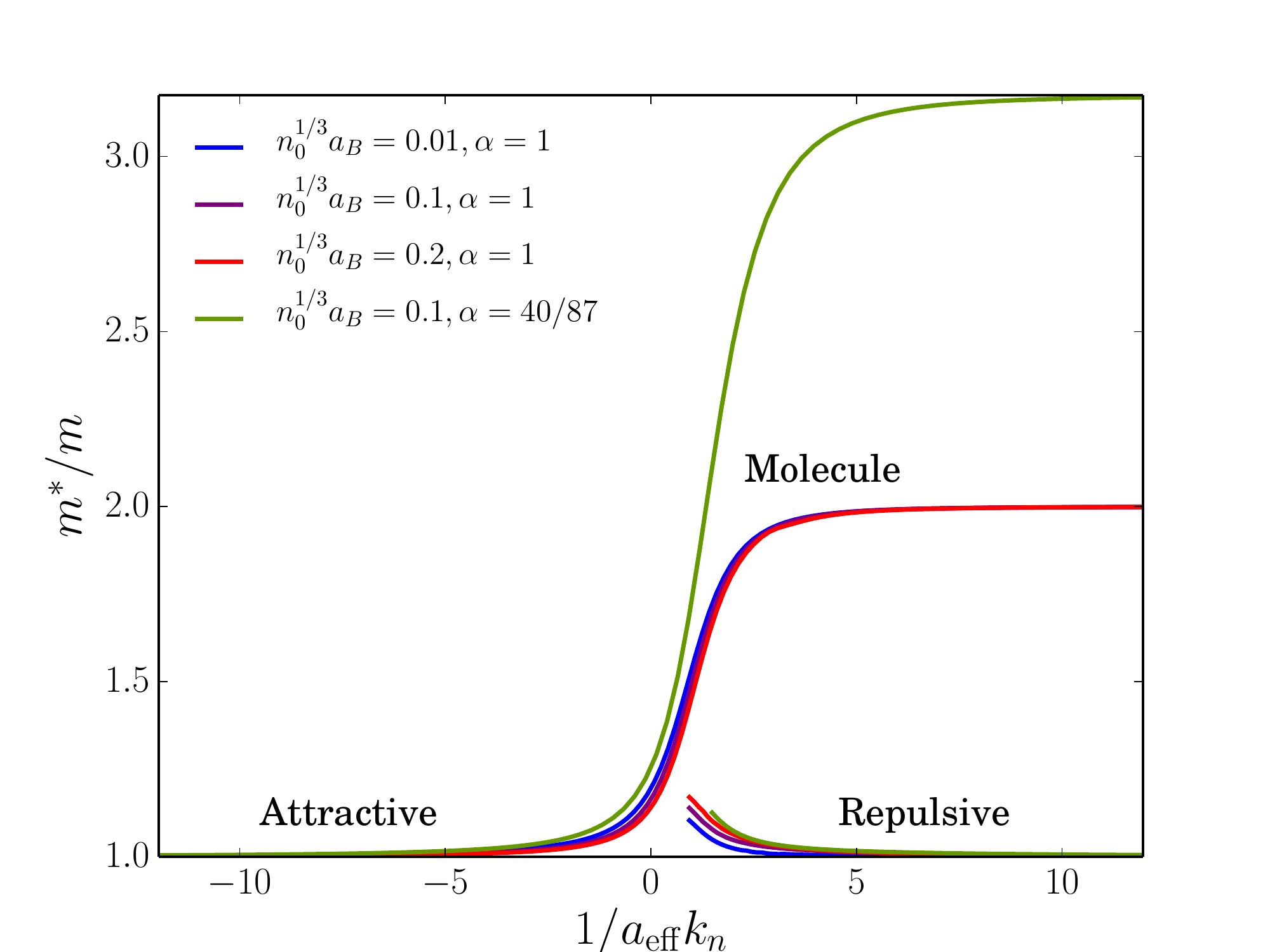}
  \caption{The effective mass for zero momentum as a function of the
    inverse Fermi-Bose interaction strength.}
  \label{fig:Mass}
\end{figure}
The corresponding quasiparticle residue and effective mass are plotted
in Figs.\ \ref{fig:Spectral}-\ref{fig:Mass}.  As for the 3D case, we
see that there are two polaronic branches: One at negative energy
$\varepsilon<0$, which is called attractive polaron, and one at
positive energy $\varepsilon>0$, which is called the repulsive
polaron.

For weak attractive interactions $1/k_na_\text{eff}\lesssim -4$, the
energy of the attractive polaron is close to the mean-field result
$gn_B$, where $n_B$ is the total density of the bosons, the residue is
$Z\simeq1$, and the effective mass is $m^*\simeq m$. As the attraction
is increased, the polaron energy decreases, but it is significantly
higher than the mean-field prediction. Contrary to the mean-field
prediction, the polaron energy is finite at unitarity
$1/k_na_\text{eff}=0$, where we find the following quasiparticle
properties: $\varepsilon/\varepsilon_n\simeq -0.54$, $Z\simeq 0.7$,
$m^*/m\simeq 1.17$ for $\alpha=1$ and $\varepsilon/\varepsilon_n\simeq
-0.75$, $Z\simeq 0.7$, $m^*/m\simeq 1.26$ for $\alpha=40/87$. These
results are \emph{universal} in the sense that they depend only weakly
on the BEC gas parameter in the range $0.01\le n_0^{1/3}a_B\le 0.2$
within the theory, as can be seen from Figs.\
\ref{fig:Energy}-\ref{fig:Mass}. This should be contrasted with the
case of a 3D Bose polaron, where a stronger dependence was found using
a the same ladder approximation~\cite{Rath2013}.  Although the
predicted universality of the polaron energy at unitarity could be an
artefact of the ladder approximation, we speculate that the dependence
on the gas parameter is suppressed in the mixed dimensional geometry,
since the impurity living in 2D affects the bosons living in 3D less.
Thus, higher order correlations neglected by the ladder approximation
might be less important in the present mixed dimensional geometry so
that the polaron has universal properties at unitarity.  This is
supported by the fact that 3-body Efimov physics is suppressed in
mixed dimensional setups as noted above~\cite{Nishida2011}. Our theory
does not predict any instability as $a_B\rightarrow 0$ in contrast to
Monte-Carlo calculations for the 3D Bose polaron, where it was
associated to the clustering of many bosons around to the
impurity~\cite{Ardila2015}. Similar effects for the 3D Bose polaron
were found in Ref.~\cite{Grusdt2017}.  Eventually, the attractive
polaron energy approaches the dimer energy $-1/2m_ra_\text{eff}^2$ on
the BEC side ($a_\text{eff}>0$) of the resonance, the residue
approaches zero, and the effective mass approaches $m^*=m+m_B$. This
reflects the fact the impurity has formed a mixed dimensional dimer
state with one boson from the BEC, in analogy with what happens for
the 3D polaron.

The repulsive polaron is well defined for weak repulsive interactions
$1/k_na_\text{eff}\gg 1$ with an energy close to the mean-field result
$gn_B$, a residue $Z\simeq1$, and an effective mass $m^*\simeq m$. As
the repulsion increases, the energy and effective mass increase,
whereas the residue decreases. We find that the polaron becomes ill
defined for strong repulsion $0 < 1/k_na_\text{eff}\lesssim 0.8$, where
the numerics cannot find the residue and effective mass due to a large
imaginary part of the self-energy.

\begin{figure}[htbp]
  \centering
  \includegraphics[width=1.07\columnwidth]{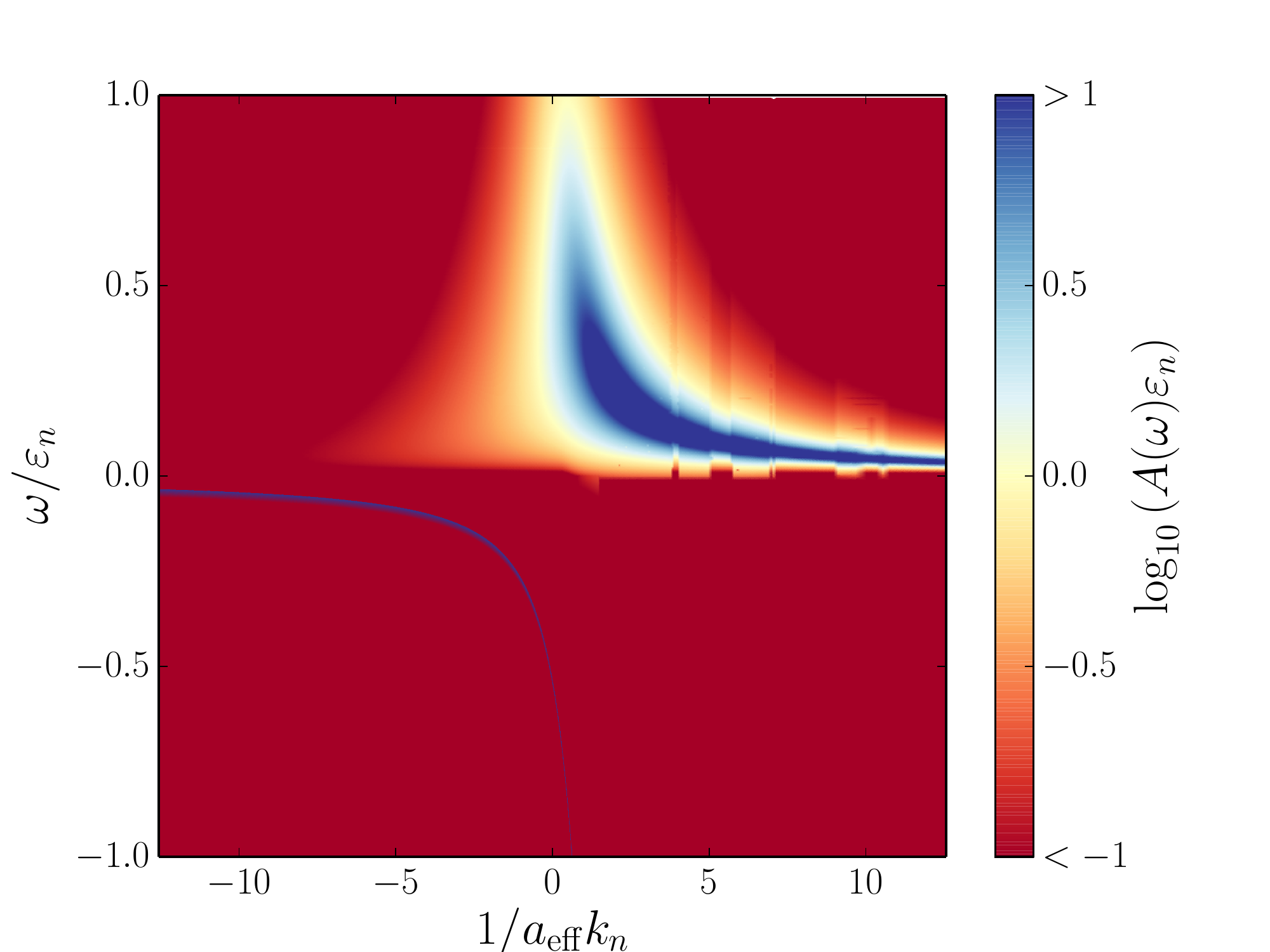}
  \caption{The spectral function of the zero momentum polaron  as a
    function of frequency and inverse Fermi-Bose interaction
    strength.}
  \label{fig:SpecFct}
\end{figure}

To investigate this further, we plot in Fig.\ \ref{fig:SpecFct} the
spectral function $A(\omega)$ of the polaron as a function of
$1/k_na_\text{eff}$ for zero temperature and a BEC gas parameter
$n_0^{1/3}a_B=0.1$.
As expected, the attractive polaron gives rise to a sharp peak with a
width given by the small imaginary part $i\delta/\varepsilon_n \leq i10^{-5}$, which
is added by hand to the frequency in the numerical calculations. We
see that there is also a continuum of spectral weight for
$\omega>0$. This continuum corresponds to states consisting of an
impurity with transverse momentum $\bp_\perp$ and a Bogoliubov mode with
momentum $\bp=(-\bp_\perp,p_z)$. Since the ladder approximation treats
the scattered impurity as a bare particle, the energy of these
continuum states are predicted to be $\omega=\bp_\perp^2/2m+E_\bp$
with a threshold at $\omega=0$. This is however not physical since the
scattered impurity also forms a polaron, and a more elaborate theory
including self-consistent impurity propagators in all diagrams would
yield a continuum starting just above the polaron quasiparticle peak
on the attractive side $a_\text{eff}<0$ of the
resonance~\cite{Rath2013}.

We see from Fig.\ \ref{fig:SpecFct} that the polaron peak on the
repulsive side $a_\text{eff}>0$ is strongly damped as the interaction
is increased towards the unitarity limit. This is because it sits
right in the middle of the continuum described above. It is due to
this strong damping that the repulsive polaron residue and effective
mass cannot be calculated for $1/k_na_\text{eff}\lesssim 0.8$ as
can be seen from Figs.\ \ref{fig:Spectral}-\ref{fig:Mass}.  This
result for the damping is however not quantitatively reliable since
the continuum is not treated in a self-consistent manner as noted
above. Also, we have not included 3-body decay of the repulsive
polaron into the dimer
state~\cite{Fedichev1996,Petrov2003,Massignan2011}. Nevertheless, we
expect the non-zero damping of the repulsive polaron predicted by the
ladder approximation to be qualitatively correct, since it does
contain the 2-body decay into a Bogoliubov mode and a scattered
impurity in an approximate way, which likely is dominant in analogy
with the 3D case~\cite{Massignan2011}.


\section{Conclusions}
In conclusion, we analysed a mixed dimensional Bose polaron, where the
impurity particle moves in a 2D plane immersed in a 3D BEC. Using a
diagrammatic ladder approximation that includes the mixed dimensional
2D-3D vacuum scattering between the impurity and the bosons
exactly, the mixed dimensional polaron was shown to exhibit the same
qualitative features as the pure 3D Bose-polaron. In particular, there is a
well defined polaron state for attractive impurity-boson interaction
that smoothly develops into a mixed dimensional dimer for strong
attraction, and there is a well defined polaron state for weak
repulsive interaction, which becomes strongly damped as the repulsion
increases. As opposed to the 3D case, our calculations predict that
the properties of the polaron are almost independent of the gas
parameter of the BEC as long as it is small so that Bogoliubov theory
applies.  It follows that the polaron has universal properties in the
unitarity limit of the impurity-boson interaction. We speculate that
higher order correlations, which could change this result, are
suppressed in the mixed dimensional geometry. The fact that we predict
well-defined quasiparticles in mixed dimensional systems indicates
that these systems should be well described by Fermi liquid theory,
which will be interesting to investigate in the future.

\begin{acknowledgments}
  G.M.B.~wishes to acknowledge the support of the Villum Foundation
  via grant VKR023163. N.J.S.L. acknowledges support by the Danish
  Council for Independent Research DFF Natural Sciences and the DFF
  Sapere Aude program.
\end{acknowledgments}

\newpage

\appendix

\onecolumngrid

\section{Appendix}

In this appendix we derive the expressions we implemented numerically
to obtain the results discussed in the paper. We will also comment on the
approximations used in order to obtain an efficient numerical code.

First we derive expressions for
$\Sigma_0(\bk_\perp,\omega)$ and $\Sigma_1(\bk_\perp,\omega)$ that
sum to be the polaron self-energy $\Sigma(\bk_\perp,\omega)$, the key
ingredient in all further computations. From the self-energy we may
directly evaluate the spectral function $A(\bk_\perp,\omega)$ from
Eq.~\eqref{eq:SpectralFct} and obtain the quasiparticle energy
$\varepsilon_{\bk_\perp}$ as the solution to
Eq.~\eqref{eq:Energy}. The quasiparticle residue $Z_{\bk_\perp}$ and
the effective mass $m^*_{\bk_\perp}$ given by
Eqs.~\eqref{eq:SpectralWeight}--\eqref{eq:Mass} require the
derivatives of the self-energy. 

\subsection{$\Sigma_0$ and its derivatives}
Computation of $\Sigma_0(\bk_\perp,\omega)$ as given in
Eq.~\eqref{eq:Diagram1} requires the pair propagator given in
Eq.~\eqref{eq:PairProp1} with $i\omega_m + \mu \rightarrow \omega +
i0^+$. In this section we consider the following:
\begin{align}
  &\Sigma_0(\boldsymbol{0}, \omega) = \frac{n_B}{g^{-1} -
    \Pi(\boldsymbol{0},\omega)}
  \label{eq:Sigma0}\\
  &\partial_\omega\Sigma_0(\boldsymbol{0}, \omega)
  = \frac{1}{n_B} (\Sigma_0(\boldsymbol{0}, \omega))^2
  \partial_\omega\Pi(\boldsymbol{0},\omega)
  \label{eq:DerivOmegaSigma0}\\
  &k_\perp^{-1} \partial_{k_\perp} \Sigma_0(\bk_\perp,\omega)|_{k_\perp=0}
  = \frac{1}{n_B} (\Sigma_0(\boldsymbol{0},\omega))^2
  k_\perp^{-1} \partial_{k_\perp} \Pi(\bk_\perp,\omega)|_{k_\perp=0}.
  \label{eq:DerivkperpSigma0}
\end{align}

We start by simplifying the pair propagator by taking the
zero-temperature limit, i.e., by setting the Bose distribution function
 $f_{\bp} = 0$. In spherical coordinates the pair propagator
becomes:
\begin{align}
  &\Pi (\bk_\perp,\omega) = \frac{2m_B}{(2\pi)^3}
  \int_0^\infty dp \, p^2 \int_0^\pi d\theta \, \sin\theta
  \int_0^{2\pi} d\phi \nonumber\\
  &\times \left[
    \left(
      \frac{p^2 + \tilde g_B}{2\tilde E_p} + \frac{1}{2}
    \right)
    \frac{1}{2m_B\omega - \tilde E_p -
      \alpha^{-1}(p^2\sin^2\theta + k_\perp^2 -
      2pk_\perp\sin\theta\cos\phi) + i0^+}
    + \frac{1}{p^2(1 + \alpha^{-1}\sin^2\theta)}
  \right]
\end{align}
with $\tilde E_p \equiv \sqrt{p^2(p^2 + 2\tilde g_B)}$ and $\tilde g_B
\equiv 2m_Bg_Bn_B$. The integral
over $\phi$ is trivial when $k_\perp = 0$, but it can be performed
also in the case $k_\perp \neq 0$. In the latter case, we let $z_0 =
(2m_B\omega - \tilde E_p - \alpha^{-1}(p^2\sin^2\theta + k^2_\perp) +
i0^+)/(2\alpha^{-1}pk_\perp\sin\theta)$ located in the upper half
complex plane. The $\phi$ integral takes the form $\int_0^{2\pi} d\phi
\, (z_0 + \cos\phi)^{-1} = 2\pi / (\sqrt{z_0-1}\sqrt{z_0+1})$, where
the complex square roots should be taken in the upper half plane. The
integral over $\theta$ can be simplified by defining $x = -
\cos\theta$ and substituting $\int_0^\pi d\theta \, \sin\theta
\rightarrow 2\int_0^1 dx$, yielding for the pair propagator:
\begin{align}
  \Pi (\bk_\perp,\omega) &= \frac{2m_B}{2\pi^2}
  \int_0^\infty dp \int_0^1 dx
 \left[
    \left(
      \frac{p^2 + \tilde g_B}{2\tilde E_p} + \frac{1}{2}
    \right)
    \frac{p^2}{\sqrt{z_+}\sqrt{z_-}}
    + \frac{1}{1 + \alpha^{-1}(1-x^2)}
  \right] \nonumber\\
  &= \frac{2m_B}{2\pi^2}
  \int_0^\infty dp
  \left[
    \left(
      \frac{p^2 + \tilde g_B}{2\tilde E_p} + \frac{1}{2}
    \right)
     \int_0^1 dx
    \frac{p^2}{
      \sqrt{z_+}\sqrt{z_-}}
    + \frac{\text{arcsinh}(\alpha^{-1/2})}{\sqrt{\alpha^{-1}(\alpha^{-1}+1)}}
  \right]
  \label{eq:PairProp_kperp}
\end{align}
with $z_\pm = 2m_B\omega - \tilde E_p - \alpha^{-1}(p\sqrt{1-x^2} \pm
k_\perp)^2 + i0^+$. In the case $k_\perp = 0$ the expression reduces
to
\begin{align}
  \Pi(\boldsymbol{0},\omega)
  &= \frac{2m_B}{2\pi^2}
  \int_0^\infty dp
  \left[
    \left(
      \frac{p^2 + \tilde g_B}{2\tilde E_p} + \frac{1}{2}
    \right)
     \int_0^1 dx
    \frac{p^2}{2m_B\omega - \tilde E_p - \alpha^{-1}p^2(1-x^2) + i0^+}
    + \frac{\text{arcsinh}(\alpha^{-1/2})}{\sqrt{\alpha^{-1}(\alpha^{-1}+1)}}
  \right] \nonumber\\
  &= \frac{2m_B}{2\pi^2}
  \int_0^\infty dp
  \left[
    \left(
      \frac{p^2 + \tilde g_B}{2\tilde E_p} + \frac{1}{2}
    \right)
     \mathcal{P} \int_0^1 dx
    \frac{p^2}{2m_B\omega - \tilde E_p - \alpha^{-1}p^2(1-x^2)}
    + \frac{\text{arcsinh}(\alpha^{-1/2})}{\sqrt{\alpha^{-1}(\alpha^{-1}+1)}}
  \right] \nonumber\\
  & - i\pi \frac{2m_B}{2\pi^2}
  \int_0^\infty dp \, p^2
    \left(
      \frac{p^2 + \tilde g_B}{2\tilde E_p} + \frac{1}{2}
    \right)
    \int_0^1 dx \,
    \delta(2m_B\omega - \tilde E_p - \alpha^{-1}p^2(1-x^2)) \; .
    \label{PProp}
\end{align}
The second equality separates the real and imaginary part of the
integral. Here $\mathcal{P}$ denotes the Cauchy principal value
integral and $\delta(x)$ is the Dirac delta function. In practice we
use the first line in Eq.~(\ref{PProp}) to calculate the real part of the integral by setting $0^+$ to a positive
number which is sufficiently small. We let $z_1 = (2m_B\omega - \tilde E_p - \alpha^{-1}p^2 +
i0^+)\alpha/p^2$ and take the $x$ integral as $\int_0^1 dx \,
(z_1+x^2)^{-1} = \text{arccot}(\sqrt{z_1})/\sqrt{z_1}$ with the complex
square root taken in the upper half plane. Hence
\begin{align}
  \Re\Pi(\boldsymbol{0},\omega)
  = \frac{2m_B}{2\pi^2}
  \int_0^\infty dp
  \left[
    \left(
      \frac{p^2 + \tilde g_B}{2\tilde E_p} + \frac{1}{2}
    \right)
    \Re\left(
      \frac{\alpha\,\text{arccot}(\sqrt{z_1})}{\sqrt{z_1}}
      \right)
    + \frac{\text{arcsinh}(\alpha^{-1/2})}{\sqrt{\alpha^{-1}(\alpha^{-1}+1)}}
  \right] \; .
\end{align}
For the imaginary part of the pair propagator, we
define a new variable $u = 1 - x^2$ and the function $u_\delta(p) =
(2m_B\omega-\tilde E_p) \alpha /p^2$ which allows us to express
\begin{align}
  \Im\Pi(\boldsymbol{0},\omega) = - \frac{2m_B}{2\pi}
  \int_0^\infty dp \,
  \left( \frac{p^2+\tilde g_B}{2\tilde E_p} + \frac{1}{2} \right)
  \int_0^1 du \, \frac{\alpha}{2\sqrt{1-u}}
  \delta(u_\delta(p) - u) \; .
\end{align}
The Dirac delta function is only non-vanishing along the $u$
integration interval for those values of $p$ where $0 < u_\delta(p) <
1$. Notice that this implies that $\Im\Pi(\boldsymbol{0},\omega) = 0$
for $\omega \leq 0$. In the case $\omega > 0$ we have to determine the
values of $p$ in the integration interval that fulfill $0 <
u_\delta(p) < 1$. Formally we may define this set as $\mathcal V = \{
p\in (0;\infty) \,\colon u_\delta(p) \in (0;1)\}$. Since the Dirac
delta function contributes only when $p\in\mathcal V$, we have
\begin{align}
  \Im\Pi(\boldsymbol{0},\omega) = - \frac{2m_B}{2\pi}
  \int_\mathcal{V} dp \,
  \left( \frac{p^2+\tilde g_B}{2\tilde E_p} + \frac{1}{2} \right)
  \frac{\alpha}{2\sqrt{1-u_\delta(p)}} \; .
\end{align}
We now prove that $\mathcal V$ is an interval. First notice that
$u_\delta(p) \rightarrow \infty$ as $p \rightarrow 0$ and $u_\delta(p)
\rightarrow -\alpha < 0$ as $p \rightarrow \infty$. Since $u_\delta$
is continous the inequality $0<u_\delta(p)<1$ is indeed fulfilled
somewhere along the $p$ integration. Second we notice from explicit
computation that the equation $du_\delta/dp = 0$ has at most one real
solution on $(0;\infty)$ which must correspond to a global
minimum. Thus $u_\delta$ decreases monotonically in the region where
$0<u_\delta(p)<1$. We conclude that $\mathcal V = (p_\text{min};
p_\text{max})$ with the end points uniquely defined by
$u_\delta(p_\text{min}) = 1$ and $u_\delta(p_\text{max}) =
0$. Explicitly we have
\begin{align}
  p_\text{min} =
  \begin{cases}
    \sqrt{\frac{-\tilde g_B\alpha^2 - \omega\alpha - \alpha\sqrt{(\tilde g_B^2 +
        \omega^2)\alpha^2 + 2\tilde g_B \omega\alpha}}{\alpha^2-1}}
    & \text{if } \alpha < 1\\
    \sqrt{\frac{\omega^2}{2(\tilde g_B + \omega)}} & \text{if } \alpha = 1\\
    \sqrt{\frac{-\tilde g_B\alpha^2 - \omega\alpha + \alpha\sqrt{(\tilde g_B^2 +
        \omega^2)\alpha^2 + 2\tilde g_B \omega\alpha}}{\alpha^2-1}}
    & \text{if } \alpha > 1
  \end{cases}
  \qquad \text{and} \qquad
  p_\text{max} =  \sqrt{-\tilde g_B + \sqrt{\tilde g_B^2 + \omega^2} }.
\end{align}
The imaginary part of the pair propagator takes the final form:
\begin{align}
  \Im\Pi(\boldsymbol{0},\omega) =
  \begin{cases}
    0 & \text{if } \omega \leq 0 \\
    - \frac{2m_B}{2\pi}
    \int_{p_\text{min}}^{p_\text{max}} dp \,
    \left( \frac{p^2+\tilde g_B}{2\tilde E_p} + \frac{1}{2} \right)
    \frac{\alpha}{2\sqrt{1-(2m_B\omega - \tilde E_p)\alpha/p^2}}
    & \text{if } \omega > 0.
  \end{cases}
\end{align}

We now turn to the derivatives of the pair propagator appearing in
Eqs.~\eqref{eq:DerivOmegaSigma0}--\eqref{eq:DerivkperpSigma0}. From
Eq.~\eqref{eq:PairProp_kperp} we find
\begin{align}
  \partial_\omega \Pi(\bk_\perp,\omega)
  = \frac{(2m_B)^2}{2\pi^2}
  \int_0^\infty dp
    \left(
      \frac{p^2 + \tilde g_B}{2\tilde E_p} + \frac{1}{2}
    \right)
     \int_0^1 dx
     \left[
    \frac{-p^2}{2(z_+)^{3/2}\sqrt{z_-}}
    + \frac{-p^2}{2(z_-)^{3/2}\sqrt{z_+}}
    \right] \; ,
\end{align}
which simplifies in the case $k_\perp=0$ using $z_- = z_+$ and
$\int_0^1 dx \, (z_1 + x^2)^{-2} = (\sqrt{z_1}/(z_1+1) +
\text{arccot}(\sqrt{z_1})/(2z_1^{3/2})$, where the complex square root
should be taken in the upper half plane. We find
\begin{align}
  \partial_\omega \Pi(\bk_\perp,\omega)
  = - \frac{(2m_B)^2}{2\pi^2}
  \int_0^\infty dp
    \left(
      \frac{p^2 + \tilde g_B}{2\tilde E_p} + \frac{1}{2}
    \right)
    \frac{\alpha^2}{2z_1^{3/2}p^2}
     \left[
       \frac{\sqrt{z_1}}{z_1+1} + \text{arccot}(\sqrt{z_1})
    \right] \; .
\end{align}
Similarly we find
\begin{align}
  k_\perp^{-1} \partial_{k_\perp} \Pi(\bk_\perp,\omega) =
  &\frac{2m_B}{2\pi^2} 
  \int_0^\infty dp 
  \left( 
    \frac{p^2 + \tilde g_B}{2\tilde E_p} + \frac{1}{2} 
  \right) \nonumber\\
  &\int_0^1 dx 
  \frac{-p^2 2\alpha^{-1}k_\perp(\alpha^{-1}(k_\perp-p\sqrt{1-x^2})(k_\perp+p\sqrt{1-x^2})
    -2m_B\omega+\tilde E_p - i0^+)}{z_+^{3/2} z_-^{3/2}} \, .
\end{align}
Notice that we may safely put $k_\perp = 0$ in the above expression
and evaluate the $x$ integral:
\begin{align}
  &k_\perp^{-1} \partial_{k_\perp} \Pi(\bk_\perp,\omega)|_{k_\perp=0} =
  \frac{2m_B}{\pi^2}
  \int_0^\infty dp
  \left(
    \frac{p^2 + \tilde g_B}{2\tilde E_p} + \frac{1}{2}
  \right)  \frac{p^2}{\alpha}
  \int_0^1 dx \,
  \frac{2m_B\omega - \tilde E_p + i0^+ + \alpha^{-1}p^2(1-x^2)}
  {(2m_B\omega - \tilde E_p + i0^+ - \alpha^{-1}p^2(1-x^2))^3} \\
  &=\frac{2m_B}{4\pi^2\alpha}
  \int_0^\infty dp
  \left(
    \frac{p^2 + \tilde g_B}{2\tilde E_p} + \frac{1}{2}
  \right)  p^2
  \left[
    \frac{3}{(2m_B\omega - \tilde E_p + i0^+ - \alpha^{-1}p^2)^2}
    + \frac{\sqrt\alpha(2m_B\omega - \tilde E_p + i0^+ + 2\alpha^{-1}p^2)
    \text{arccot}(z_1)}{p (2m_B\omega - \tilde E_p + i0^+ - \alpha^{-1}p^2)^{5/2}}
  \right] \; .
\end{align}
This concludes the derivations of the numerical integrals we
implemented in order to compute
Eqs.~\eqref{eq:Sigma0}--\eqref{eq:DerivkperpSigma0}.

\subsection{$\Sigma_1$ and its derivatives}
In this section we shall compute $\Sigma_1(\boldsymbol{0},\omega)$ and
the derivatives $\partial_\omega\Sigma_1(\boldsymbol{0}, \omega)$ and
$k_\perp^{-1} \partial_{k_\perp}
\Sigma_1(\bk_\perp,\omega)|_{k_\perp=0}$. From
Eq.~\eqref{eq:Diagram2}, with $i\omega_n + \mu \rightarrow \omega + i0^+$,
we have
\begin{align}
  \Sigma_1(\bk_\perp, \omega) &=
  -T \sum_\nu \int \frac{d^3p}{(2\pi)^3}
  \left.
  \left(
    \frac{u_\bp^2}{i\omega_\nu - E_\bp}
    - \frac{v_\bp^2}{i\omega_\nu + E_\bp}
  \right)
  \frac{1}{g^{-1} - \Pi(\bp_\perp + \bk_\perp,i\omega_n + i\omega_\nu)}
  \right|_{i\omega_n + \mu \rightarrow \omega + i0^+}\\
  &= \int \frac{d^3p}{(2\pi)^3}
    \frac{v_\bp^2}{g^{-1} - \Pi(\bp_\perp+\bk_\perp,\omega - E_\bp)} \; ,
  \label{eq:Sigma1}
\end{align}
where the last line is found from performing the sum over Matsubara
frequencies and letting the temperature $T \rightarrow 0$. Here we
exploit that the chemical potential $\mu$ is minus infinity such that
any poles and branch cuts of the $\mathcal{T}$-matrix are pushed to
infinity.

We have to simplify the expression above in order to get a numerical
feasible implementation. Therefore we approximate the pair propagator
by that for a non-interaction Bose gas at zero temperature $\Pi_{\rm vac}$. This
amounts to setting $g_B = 0$ and $f_\bp = 0$ in
Eq.~\eqref{eq:PairProp1}. We now show that it reduces to the
expression in Eq.~\eqref{eq:PairpropagatorVacuumAna}. Notice that
these approximations are only applied to the pair propagator, as
setting the temperature to zero in the entire expression for
$\Sigma_1$ would make it vanish, and this we are definitely not
interested in. The pair propagator takes the form
\begin{align}
  \Pi_{\rm vac}(\bp_\perp,\omega - E_\bp)
  = \int\frac{d^3\tilde p}{(2\pi)^3}
  \left[
    \frac{1}{\omega - E_\bp - \tilde p^2/2m_B 
    - (\bk_\perp + \bp_\perp + \tilde\bp_\perp)^2/2m + i0^+}
  + \frac{2m_B}{\tilde p^2 + \tilde p_\perp^2/\alpha}
  \right] \; .
\end{align}
We shift $\tilde \bp$ in the first term in the integrand by adding the
constant vector $\bp \, m_B/M$. Then we scale $\tilde \bp_\perp$ in the
entire integrand by the factor $\sqrt{m/M}$ such that the integral,
with $z' = 2m_B(\omega - E_\bp - (\bp_\perp + \bk_\perp)^2/2M +
i0^+)$, becomes
\begin{align}
  \Pi_{\rm vac}(\bp_\perp + \bk_\perp,\omega - E_\bp)
  &= \frac{2m_B}{1+\alpha^{-1}}
  \int\frac{d^3\tilde p}{(2\pi)^3}
  \left[
    \frac{1}{z' - \tilde p^2}
  + \frac{1}{\tilde p^2}
  \right] \\
  &= \frac{2m_B}{2\pi^2(1+\alpha^{-1})} \int_0^\infty d\tilde p \,
  \frac{z'}
  {z' - \tilde p^2} \\
  &= - \frac{2m_B z'}{4\pi^2(1+\alpha^{-1})}
  \int_{-\infty}^\infty d\tilde p \,
  \frac{1}{(\tilde p + \sqrt{z'}) (\tilde p - \sqrt{z'})} \; .
\end{align}
Noting that the pole at $\sqrt{z'}$ is located in the upper half
complex plane, we perform the contour integral around the pole yielding
$\int_{-\infty}^\infty d\tilde p \, [(\tilde p + \sqrt{z'})(\tilde p -
\sqrt{z'})]^{-1} = i\pi/\sqrt{z'}$. The pair propagator
simplifies to
\begin{align}
  \Pi_{\rm vac}(\bp_\perp + \bk_\perp,\omega - E_\bp) =
  \begin{cases}
    -i \frac{\sqrt{m_B}m_r}{\sqrt{2} \pi}
    \sqrt{|\omega - E_\bp - (\bp_\perp + \bk_\perp)^2/2M|}
    & \qquad \text{if }\; \omega - E_\bp - (\bp_\perp + \bk_\perp)^2/2M \geq 0\\[1em]
    \frac{\sqrt{m_B}m_r}{\sqrt{2} \pi}
    \sqrt{|\omega - E_\bp - (\bp_\perp + \bk_\perp)^2/2M|}
    & \qquad \text{if }\; \omega - E_\bp - (\bp_\perp + \bk_\perp)^2/2M < 0
  \end{cases}
  \label{eq:PairPropSimpelkperp}
\end{align}
which is equivalent to the expression in
Eq.~\eqref{eq:PairpropagatorVacuumAna}.

Returning to $\Sigma_1$ from Eq.~\eqref{eq:Sigma1} we go to spherical
coordinates $(p,\theta,\phi)$ and substitute $x=-\cos\theta$:
\begin{equation}
  \Sigma_1(\bk_\perp, \omega) =
  \frac{1}{2\pi^3}
  \int_ 0^\infty dp \, \int_0^1 dx \, \int_0^\pi d\phi
  \frac{p^2 \, v_\bp^2}{g^{-1} - e^{i\psi_-}\frac{\sqrt{m_B} m_r}{\sqrt 2 \pi}
    \sqrt{|\omega - E_\bp - (k_\perp^2 + p^2(1-x^2) + 2k_\perp p\sqrt{1-x^2}\cos\phi)/2M|}}
  \; ,
  \label{eq:Sigma1kperp}
\end{equation}
where $e^{i\psi_\pm} \in \{1, -i \}$ are integration variable
dependent phase factors given according to
Eq.~\eqref{eq:PairPropSimpelkperp}. In the case $k_\perp=0$ the $\phi$
integration is trivial, and we get
\begin{equation}
  \Sigma_1(\boldsymbol{0}, \omega) =
  \frac{1}{2\pi^2}
  \int_ 0^\infty dp \, \int_0^1 dx \,
    \frac{p^2 \, v_\bp^2}{g^{-1} - e^{i\psi_-}\frac{\sqrt{m_B} m_r}{\sqrt 2 \pi}
    \sqrt{|\omega - E_\bp - p^2(1-x^2)/2M|}} \; .
\end{equation}

The derivate of $\Sigma_1(\boldsymbol{0},\omega)$ with respect to
$\omega$ is straight-forward to compute:
\begin{equation}
  \partial_\omega \Sigma_1(\boldsymbol{0}, \omega) =
  -\frac{\sqrt{m_B}m_r}{4\sqrt 2 \pi^3}
  \int_ 0^\infty dp \, \int_0^1 dx \,
    \frac{p^2 \, v_\bp^2 \, e^{i\psi_-}(\omega - E_\bp - p^2(1-x^2)/2M)}
    {|\omega - E_\bp - p^2(1-x^2)/2M|^{3/2}
      \left(
        g^{-1} - e^{i\psi_-}\frac{\sqrt{m_B} m_r}{\sqrt 2 \pi}
        \sqrt{|\omega - E_\bp - p^2(1-x^2)/2M|}
      \right)^2} \; .
\end{equation}
Finally we compute the limit of $k_\perp^{-1} \partial_{k_\perp}
\Sigma_1(\bk_\perp,\omega)$ when $k_\perp \rightarrow 0$ from
Eq.~\eqref{eq:Sigma1kperp}. We notice that the integrand of
$k_\perp^{-1} \partial_{k_\perp} \Sigma_1(\bk_\perp,\omega)$ consists of two term when
$k_\perp$ is small. One of the terms is
proportional to $k_\perp^{-1}\cos\phi$ and the integral over $\phi$
vanishes. The other term is constant with respect to $\phi$ and
$k_\perp$, and the $\phi$ integral just yields a factor of $\pi$. All taken together, we find that $k_\perp^{-1} \partial_{k_\perp}
\Sigma_1(\bk_\perp, \omega)|_{k_\perp=0} = -M^{-1} \partial_\omega
\Sigma_1(\boldsymbol{0}, \omega)$, and so we do not have to implement this
formula separately.

\twocolumngrid

\bibliographystyle{apsrev4-1}
\bibliography{RefMixedPolaron}

\end{document}